\documentclass{kapedbk} 

\usepackage[dvips]{graphicx}

\chaptersection 

\kluwerbib 

\begin{document}

\articletitle[Black hole accretion]
{Theory of disk accretion onto \\ supermassive black holes}

\rhead{Black hole accretion theory}

\author{Philip J. Armitage}

\affil{JILA and the Department of Astrophysical and Planetary Sciences, 
University of Colorado, Boulder CO~80309-0440, USA
\footnote{Partially supported by {\em NASA} and {\em PPARC}.}
}
\email{pja@jilau1.colorado.edu}

\begin{abstract}
Accretion onto supermassive black holes produces both the 
dramatic phenomena associated with active galactic nuclei
and the underwhelming displays seen in the Galactic Center 
and most other nearby galaxies. I review selected aspects of 
the current theoretical 
understanding of black hole accretion, emphasizing the role of 
magnetohydrodynamic turbulence and gravitational instabilities 
in driving the actual accretion 
and the importance of the efficacy of cooling in determining the 
structure and observational appearance of the accretion flow. 
Ongoing investigations into the dynamics of the plunging region, 
the origin of variability in the accretion process, and the evolution of 
warped, twisted, or eccentric disks are summarized.
\end{abstract}


\section{Introduction}

Most galactic nuclei are now believed to harbor supermassive black holes.  
These black holes are all accreting gas---at a minimum from the 
interstellar medium proximate to the 
event horizon, and in some cases from dense disks of gas in 
orbit around the hole. The observational signatures of this 
accretion, however, differ dramatically from galaxy to galaxy. 
The black hole in the Galactic Center, for example, has an 
X-ray luminosity in quiescence of only $L_X \simeq 2 \times 10^{33} \ 
{\rm ergs \ s}^{-1}$, reaching $L_X \sim 10^{35} \ 
{\rm ergs \ s}^{-1}$ during flaring states (\cite{pja:baganoff01}, 2003; 
\cite{pja:goldwurm03}). Many nearby 
elliptical galaxies also show nuclear emission that is much 
weaker than might be expected on the basis of simple 
estimates of the black hole accretion rate (\cite{pja:fabian95}; 
\cite{pja:dimatteo00}; \cite{pja:loewenstein01}). In stark contrast 
to these feeble displays, accretion onto black holes in quasars 
powers the most luminous steady sources in the universe. At a 
minimum, a theory of accretion needs to account for this 
dichotomy, and to explain at least some of the many 
secondary phenomena (jets, outflows, variability, etc.) 
associated with Active Galactic Nuclei (AGN). More ambitiously, 
one might hope to understand the role of accretion in actually 
forming supermassive black holes. Black hole formation is  
a difficult theoretical problem which, although currently 
untroubled by direct observations, is receiving increasing attention.

In this chapter, I discuss selected aspects of the theory of 
accretion onto supermassive black holes, with an emphasis 
on the physical processes that drive accretion and 
determine the qualitative properties of the flow. 
In \S\ref{secpjatransport}, 
I outline the mechanisms which can lead to angular momentum 
transport within an accretion flow, thereby allowing 
rotationally supported gas to 
flow inward and liberate gravitational potential energy. 
Close to the black hole, turbulence driven by magnetohydrodynamic 
instabilities is probably the dominant mechanism for 
transport. Further out, other processes, such as gravitational 
instabilities, are likely to assume that role. Angular momentum 
transport, however, is only part of the story.  Although 
central to our understanding of accretion, knowledge  
of its origin does not suffice to explain why the Galactic 
Center looks nothing like a powerful AGN. For that we 
need to consider the distinction between geometrically 
thin accretion disks, in which the gas can radiate 
efficiently and cool to sub-virial temperatures, and 
hot thick disks, which are radiatively inefficient and 
vulnerable to rapid mass loss. These questions are 
addressed in \S\ref{secpjaaccrate}. 
Subsequent sections examine the status of several 
open questions in the study of black hole accretion, including 
the dynamics of gas executing its final plunge into the hole, 
the origin of variability, and the evolution of disks that are 
warped or eccentric.

\section{Angular Momentum Transport}
\label{secpjatransport}

In almost all circumstances, gas in the nuclear regions of galaxies 
has far too much angular momentum to be swallowed 
directly by the black hole. Understanding the mechanisms that can 
lead to angular momentum transport in rotating flows is therefore 
the central problem in the theory of accretion onto supermassive 
black holes. At scales of $10-100$~pc or larger, this is a problem 
in galactic dynamics. Following galactic mergers, for example, 
gravitational torques from bars and 
other transient structures can efficiently funnel gas into the 
nuclear regions (Shlosman, Frank, \& Begelman 1989;
\cite{pja:barnes91}; \cite{pja:hernquist95}). 
Once within the black hole's sphere of influence, however, which 
for a galaxy of central velocity dispersion $\sigma$ and black 
hole mass $M_\bullet$ extends out to
\begin{equation} 
r_{\bullet} = { {GM_\bullet} \over \sigma^2 },
\end{equation} 
these galactic mechanisms become less efficient. Within the sphere 
of influence, which reaches $\approx 10$~pc for a black hole of 
mass $M_\bullet = 10^8 \ {\rm M}_\odot$, 
the black hole's gravity overwhelms that of the host galaxy and 
dominates the potential, unless a comparably large mass of gas 
has managed to accumulate at such small radii. Timescale arguments 
suggest that gas at $r_\bullet$ probably {\em cannot} trickle 
down through an accretion disk all the way to the vicinity of the 
black hole (e.g., \cite{pja:isaac90}). Disks probably do form, 
however, at smaller radii of order 0.1~pc, where maser emission 
in the nucleus of NCG~4258 has an unmistakably disk-like 
geometry (\cite{pja:miyoshi95}). At such radii, the specific 
angular momentum of a Keplerian disk exceeds that needed for 
direct capture by the black hole by a factor of order $10^2$. 
If that gas is to accrete, 99\% of the angular momentum 
must either be redistributed within the disk---angular momentum 
transport---or lost entirely from the system. We will consider 
the known mechanisms which can accomplish this feat. 

In the inner disk, {\em turbulence driven by magnetic 
instabilities} is a potent source of angular 
momentum transport. Although magnetohydrodynamic (MHD) disk turbulence 
has long been implicated in disk angular momentum transport and is 
mentioned prominently by Shakura \& Sunyaev (1973), current 
confidence in this conclusion rests on two more recent developments. 
First, Balbus \& Hawley (1991) demonstrated that the introduction 
of a weak magnetic field renders accretion flows linearly unstable 
to a powerful local instability. Subsequent numerical simulations 
(\cite{pja:hawley95}; \cite{pja:brandenburg95}; 
\cite{pja:matsumoto95}; \cite{pja:stone96})
showed that the instability rapidly develops into sustained 
turbulence, which transports angular momentum outward at a rate 
that is consistent with observational constraints derived from 
studies of accretion in mass transfer binaries 
(e.g., \cite{pja:pringle86}; \cite{pja:cannizzo93}; \cite{pja:osaki96}; 
\cite{pja:hameury98}). For a recent review of the role of MHD 
turbulence in accretion disks, see Balbus (2003). 

In the absence of magnetic fields, a differentially rotating 
disk with angular velocity $\Omega(r)$ is linearly stable to 
axisymmetric perturbations according to the Rayleigh criterion if
\begin{equation}
 { {\rm d} \over {{\rm d}r} } \left( r^2 \Omega \right) > 0 \,,
\end{equation}
i.e., if the specific angular momentum $l(r)$ of the flow is an 
increasing function of radius. For a geometrically thin disk
orbiting a point mass $M$, the angular velocity is Keplerian,
\begin{equation} 
 \Omega = \sqrt{ {GM} \over r^3} \,,
\end{equation} 
the specific angular momentum $l(r) \propto \sqrt{r}$, and 
the disk is hydrodynamically stable. Numerical simulations 
support this conclusion (\cite{pja:balbus96}). Although real 
disks---especially geometrically thick flows with significant 
pressure support---can have angular velocity profiles 
that differ from the simple Keplerian 
form, they too are invariably stable by the Rayleigh criterion.

Matters are drastically different if the disk contains a  
magnetic field. Analytic studies have shown that 
a weak magnetic field destabilizes astrophysical disks, 
provided that
\begin{equation}
 { {{\rm d} \Omega^2} \over {{\rm d} \ln r} } < 0 \,,
\end{equation}
a condition which is almost always satisfied in real disks. 
This {\em magnetorotational} instability (MRI) exists
regardless of the initial magnetic field configuration 
(\cite{pja:velikhov59}; \cite{pja:chandra60}; 
Balbus \& Hawley 1991, 1992; \cite{pja:terquem95}; 
\cite{pja:ogilvie96}; \cite{pja:curry96}), though the growth rates 
vary depending upon the magnetic field geometry, being fastest 
for vertical fields. 

Proving the existence of the MRI  
in the general case requires a moderately involved calculation, 
which can be found in the comprehensive review by Balbus \& 
Hawley (1998). Analyses of much simpler systems, however, 
reveal most of the important physics. Here I follow closely 
the treatment of Balbus \& Hawley (2000). Consider a fluid 
element orbiting in a disk with a central gravitational 
potential $\Phi(r)$. We will assume that pressure forces 
are negligible, as they will be, provided that perturbations 
to the disk velocity field are highly subsonic. In 
cylindrical polar coordinates $(r,z,\phi)$, the equations of 
motion then read
\begin{eqnarray} 
 \ddot{r} - r {\dot{\phi}}^2 &=& - { {\partial \Phi} \over {\partial r} } + f_r \nonumber \,, \\
 r \ddot{\phi} + 2 \dot{r} \dot{\phi} & = & f_\phi \,,
\label{pja:eq_motion} 
\end{eqnarray}
where the dots denote derivatives with respect to time, and $f_r$ and $f_\phi$ 
are forces that we will specify shortly. We now concentrate our
attention on a small patch of the disk at radius $r_0$ that is corotating 
with the overall orbital motion at angular velocity $\Omega$. We 
define a local Cartesian coordinate system $(x,y)$ via
\begin{eqnarray}
 r & = & r_0 + x \,, \\
 \phi & = & \Omega t + { y \over r_0 } \,,
\end{eqnarray}
and substitute these expressions into equation (\ref{pja:eq_motion}) above. 
Discarding quadratic terms, the result is
\begin{eqnarray} 
 \ddot{x} - 2 \Omega \dot{y} & = & - x { { {\rm d} \Omega^2} \over 
 { {\rm d} \ln r} } + f_x \nonumber \,, \\
 \ddot{y} + 2 \Omega \dot{x} & = & f_y \,.
\label{pja:eq_hill}
\end{eqnarray}  
These equations describe the epicyclic motion of pressureless fluid 
perturbed from an initially circular orbit.

If the disk contains a weak vertical magnetic field, perturbations 
to the fluid in the plane of the disk will be opposed by magnetic 
tension forces generated by the bending of the field lines. Considering 
in-plane perturbations varying with height $z$ and time $t$ as 
$e^{i (\omega t - k z)}$, the magnetic tension force is 
${\bf f} = -(kv_A)^2 {\bf s}$, 
where $\bf s$ is the displacement vector and $v_A = \sqrt{ B_z^2 / 4 \pi \rho}$ 
is the Alfv\'en speed. Using this expression for $f_x$ and $f_y$, and 
assuming the $e^{i \omega t}$ time dependence, equation (\ref{pja:eq_hill}) 
becomes
\begin{eqnarray} 
 -\omega^2 x - 2 i \omega \Omega y & = & - x { { {\rm d} \Omega^2} \over 
 { {\rm d} \ln r} } - (k v_A)^2 x \,, \\
 -\omega^2 y + 2 i \omega \Omega x & = & - (k v_A)^2 y \,. 
\end{eqnarray} 
Combining these equations yields a dispersion relation that is 
a quadratic in $\omega^2$,
\begin{eqnarray} 
 \omega^4 - \omega^2 \left[ { { {\rm d} \Omega^2} \over 
 { {\rm d} \ln r} } + 4 \Omega^2 + 2 \left( k v_A \right)^2 \right] \nonumber \\
 + \left( k v_A \right)^2 \left[ \left( k v_A \right)^2 + { { {\rm d} \Omega^2} \over 
 { {\rm d} \ln r} } \right] = 0 \,.
\end{eqnarray} 
The system is unstable if $\omega^2 < 0$, which occurs when the third 
term in the dispersion relation is negative. Instability therefore occurs 
if
\begin{equation}
 \left( k v_A \right)^2 + { { {\rm d} \Omega^2} \over 
 { {\rm d} \ln r} } < 0 \,.
\end{equation}
For a sufficiently weak field ($v_A \rightarrow 0$), or for 
long enough wavelength perturbations ($k \rightarrow 0$), 
we then obtain the aforementioned criterion for a disk 
to be unstable to the MRI, namely
\begin{equation}
 { {{\rm d} \Omega^2} \over {{\rm d} \ln r} } < 0 \,.
\end{equation}
In a real disk, the longest wavelength perturbation in 
the vertical direction will be of order the scale 
height $h$ of the disk. There will therefore be instability, 
provided that the magnetic field is weaker than some 
threshold value $B_{\rm max}$.   

The physical origin for this instability is fairly straightforward. 
Adopting the same configuration that we have just analyzed, consider 
the effect of perturbing a weak vertical magnetic field 
threading an otherwise uniform disk. If the field remains 
frozen into the plasma, then field lines which connect 
adjacent annuli in the disk will be sheared by the 
differential rotation of the disk into a trailing 
spiral pattern. Provided that the field is weak 
enough, magnetic tension is inadequate to snap the 
field lines back to the vertical. What tension there 
is, however, acts to reduce the angular momentum of 
the inner fluid element and boost that of the 
outer fluid element, providing angular momentum 
transport in the outward sense that is required to drive accretion.

The pervasive nature of the MRI, taken together with its 
rapid linear growth rate---$\omega_{\rm max}$ can be as large 
as $3\Omega / 4$, meaning that growth of magnetic field 
energy occurs on a dynamical timescale---implies that 
well-ionized astrophysical disks will be unstable. Analytic 
studies are unable, however, to investigate the most 
interesting consequence of the instabilities, namely, what happens 
when the MRI reaches a saturated state and turbulence has 
developed throughout the flow? Quantities we would like 
to determine in this state include the $r\phi$ 
component of the stress tensor,
\begin{equation}
 W_{r \phi} = \langle v_r dv_\phi - v_{Ar} v_{A\phi} \rangle \,,
\end{equation} 
which includes both
fluid (Reynolds) and magnetic (Maxwell) stresses.
In this expression, $d v_\phi = v_\phi - r \Omega$ is the 
fluctuation in the azimuthal velocity, $v_{Ar}$ and 
$v_{A\phi}$ are Alfv\'en velocities in the radial and 
azimuthal directions, respectively, and the angle brackets 
denote a density weighted average over height. If 
$W_{r \phi}$ can be measured, in practice from a 
numerical MHD simulation, then it can be used to 
estimate the $\alpha$ parameter introduced by 
Shakura \& Sunyaev (1973). They suggested the scaling
\begin{equation} 
 W_{r \phi} = \alpha c_s^2 \,,
\label{pja:eq_ss} 
\end{equation} 
where $c_s$ is the sound speed and $\alpha$ is a dimensionless 
constant or parameter that measures the 
efficiency of angular momentum transport in disks.

Numerical simulations of the non-linear development of the 
MRI have now been performed in both local and global geometries. 
Local simulations (\cite{pja:hawley95}, 1996; \cite{pja:brandenburg95}; 
\cite{pja:matsumoto95}; \cite{pja:stone96}; \cite{pja:miller00}) follow 
the evolution of the MRI in a small patch of disk, with periodic boundary 
conditions in $\phi$ and (in a modified form) $r$. Global simulations 
(\cite{pja:armitage98}; \cite{pja:hawley00}) have the advantage of 
being able to study large scale magnetic fields and are essential 
for the investigation of geometrically thick ($h \sim r$) flows 
(e.g., \cite{pja:hawley01}). All such simulations have known 
limitations. The physical separation between the largest relevant 
scale (that of the whole disk, or the wavelength of the most 
important MRI modes) and the dissipative scale (either viscous 
or resistive) is almost always too large to resolve numerically.  
Moreover, many calculations rely on purely numerical 
effects at the grid scale to provide dissipation. Discussions 
of the effect of these limitations can be found, for example, in 
Balbus \& Hawley (1998) or Schekochihin et al.\ (2004). 
Nevertheless, it is encouraging that a variety of simulations, 
using different numerical techniques, agree in 
several critical aspects. The MRI rapidly leads---on a 
timescale of just a few orbital periods---to a state of 
MHD disk turbulence in which angular momentum is transported 
outward. This is sustained, despite the presence of dissipation 
(often at unphysically large scales) within the numerical 
codes. The magnetic field energy remains smaller than 
the thermal energy in the disk. Finally, the Maxwell stress 
dominates by a large margin over the Reynolds stress, though the 
latter also provides some outward transport. Values of $\alpha$ 
between $10^{-2}$ and $10^{-1}$ are representative of the 
bottom line of most simulations.

The maintenance of a disk magnetic field, despite the presence of 
dissipation, implies that turbulence driven by the MRI sustains 
a disk dynamo, just as fluid motions in stars and planets 
generate their own self-sustaining magnetic fields 
(\cite{pja:parker55}; \cite{pja:proctor94}; \cite{pja:glatzmaier95}). 
The accretion disk case, however, is interestingly different 
from these better-observed systems. In a disk (at least 
according to current belief) there would be no turbulent 
motions in the absence of a magnetic field, which rather 
must generate the turbulent velocity field needed for its 
own amplification. This bootstrapping quality  
makes it harder to apply knowledge gleaned from the 
long history of work on solar and planetary dynamos 
to the disk case.

How do the values of $\alpha$ obtained from numerical 
simulations compare with those inferred 
from observations? The best constraints come from modeling 
the episodic accretion events seen in the dwarf novae 
subclass of cataclysmic variables. In the widely accepted 
thermal disk instability model for dwarf nova outbursts 
(\cite{pja:meyer81}; \cite{pja:mineshige83}; \cite{pja:faulkner83}), 
the eruptive behavior results from the inability of the disk 
to reach a steady-state at radii where hydrogen is 
partially ionized. Under these conditions, there is 
no thermally stable vertical structure for a range of 
accretion rates, resulting in a limit cycle in which 
the disk alternates between a cool mass-accumulating 
phase and a hot draining phase.   
The timescales of the outbursts are directly related to the 
values of $\alpha$ in the outburst and quiescent states. 
Different authors (e.g., Pringle et al.\ 1986;
\cite{pja:cannizzo93}; \cite{pja:osaki96}; \cite{pja:hameury98}) 
derive typical values of $\alpha$ in the outburst state of $\sim 0.1$, 
with quiescent values a factor of a few lower. This is in 
reasonably encouraging agreement with the values predicted 
from MHD simulations. Moreover, as I will discuss later in the 
context of AGN disks, the 
{\em difference} in $\alpha$ between the outburst and quiescent 
states could be caused by a suppression of the MRI when 
the disk is cool and the magnetic field is imperfectly 
coupled to the gas (Gammie \& Menou 1998).

Of course, in many circumstances of interest, it is either 
inconvenient or unfeasible to determine the structure and 
evolution of disks via three-dimensional MHD simulations. 
In particular, the evolution of AGN disks across the full 
range of radii spanned by real disks cannot be simulated 
directly. This raises the following questions: under what 
circumstances are vertically-integrated one-dimensional disk 
models, or two-dimensional simulations using a parameterized 
viscosity, good approximations to the real disk physics?

For geometrically thin disks ($h/r \ll 1$), the standard 
time-dependent one-dimensional treatment divides the problem 
into two parts (e.g., \cite{pja:frank02}). First, one 
solves for the vertical structure of an isolated annulus 
of the disk using methods and equations 
analogous to those of stellar structure. The vertical structure 
calculation yields the 
functional dependence of the kinematic viscosity $\nu$ on 
the surface density $\Sigma$, angular velocity $\Omega$, 
and $\alpha$. Armed with this knowledge, we can then 
solve a one-dimensional diffusion equation for the time-dependence 
of the disk,
\begin{equation}
 { {\partial \Sigma} \over {\partial t} } = {3 \over r} 
 { \partial \over {\partial r} } \left[ r^{1/2} 
 { \partial \over {\partial r} } \left( \nu \Sigma r^{1/2} \right) 
 \right] \,,
\label{pja:eq_diffuse} 
\end{equation} 
subject to appropriate boundary conditions at the inner and 
outer edges of the disk. If thermal fronts are present (or 
develop) within the disk, this equation for $\Sigma$ needs 
to be supplemented with an explicit equation for the 
time-dependence of the central temperature $T_c$ (e.g., 
\cite{pja:cannizzo93}).

Provided that we restrict our attention to the axisymmetric evolution 
of geometrically thin disks, one-dimensional disk models constructed 
in this manner probably provide a reasonable approximation to 
the evolution of MRI-active disks. At the most basic level, the 
MRI is a local instability, and, as such, is in principle amenable 
to an $\alpha$-type description (\cite{pja:bp99}). For the vertical 
structure part of the calculation, the main question is whether the 
distribution of the stress with height follows the simple 
scalings (stress proportional to pressure) normally employed 
as a generalization of the $\alpha$ {\em ansatz} (Eq.~\ref{pja:eq_ss}). 
In fact, numerical simulations of the MRI suggest that the stress 
may be roughly constant within the 
body of the disk (up to $\approx 2h$), with a significant 
fraction (perhaps a quarter) of the magnetic energy 
escaping buoyantly from the disk to form a hot, strongly 
magnetized corona (Miller \& Stone 2000). These simulation 
results differ from the simplest assumptions 
used in modeling vertical structure, but they do not 
invalidate the basic approach. Indeed, it ought 
to be possible to compute vertical structure models---and vital 
observational properties of the disk, such as the local 
spectrum (e.g., \cite{pja:hubeny97}; \cite{pja:hubeny01})---using 
the computational results as a guide to where in the 
disk dissipation of energy occurs. This has not yet been done.

%
%
\begin{figure}
\includegraphics[width=\textwidth]{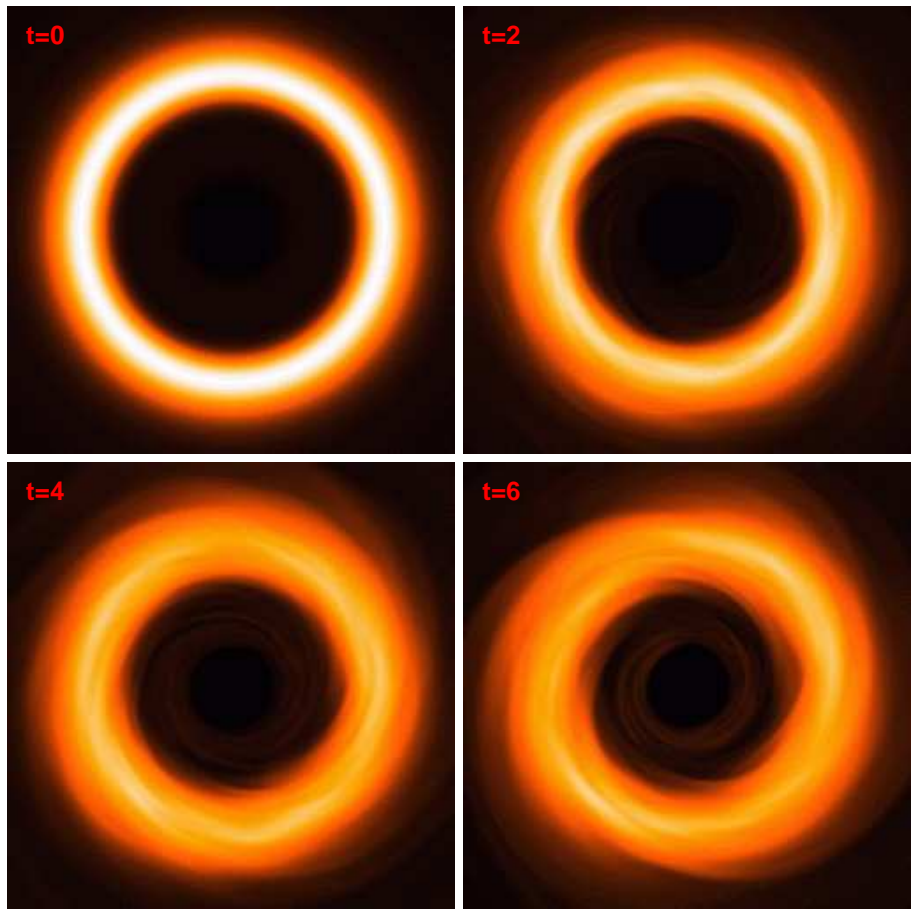}
\caption{Evolution of a narrow ring of gas under the influence
of angular momentum transport, computed using a three-dimensional
global MHD simulation (Armitage, unpublished). Panels show the
surface density $\Sigma(r,\phi)$ at different times $t$, where
$t$ is measured in units of the orbital period at the initial
center of the ring. The simulation is isothermal, with
$c_s / v_\phi = 0.2$ at $r = r_0$, the initial center of the ring.
The MRI was triggered by threading an initially stable and axisymmetric
ring of gas with a weak vertical magnetic field. A cylindrical
disk approximation was used (i.e., no vertical stratification),
and the evolution was computed using the ZEUS code (Stone \&
Norman 1992) with $200 \times 250 \times 40$ grid points in $r$,
$\phi$, and $z$, respectively. As in the one-dimensional viscous
disk solution to this problem, given
by Lynden-Bell \& Pringle (1974), the ring spreads out, with the
mass flowing inward, while a fraction of the mass at large radii
absorbs the angular momentum. In addition, however, there is
obvious non-axisymmetric structure in the form of trailing
spiral waves and relatively long-lived fluctuations in the
magnitude of the derived stress.}
\label{pja:fig1}
\end{figure}

%
%
\begin{figure}[tbh]
\includegraphics[width=\textwidth]{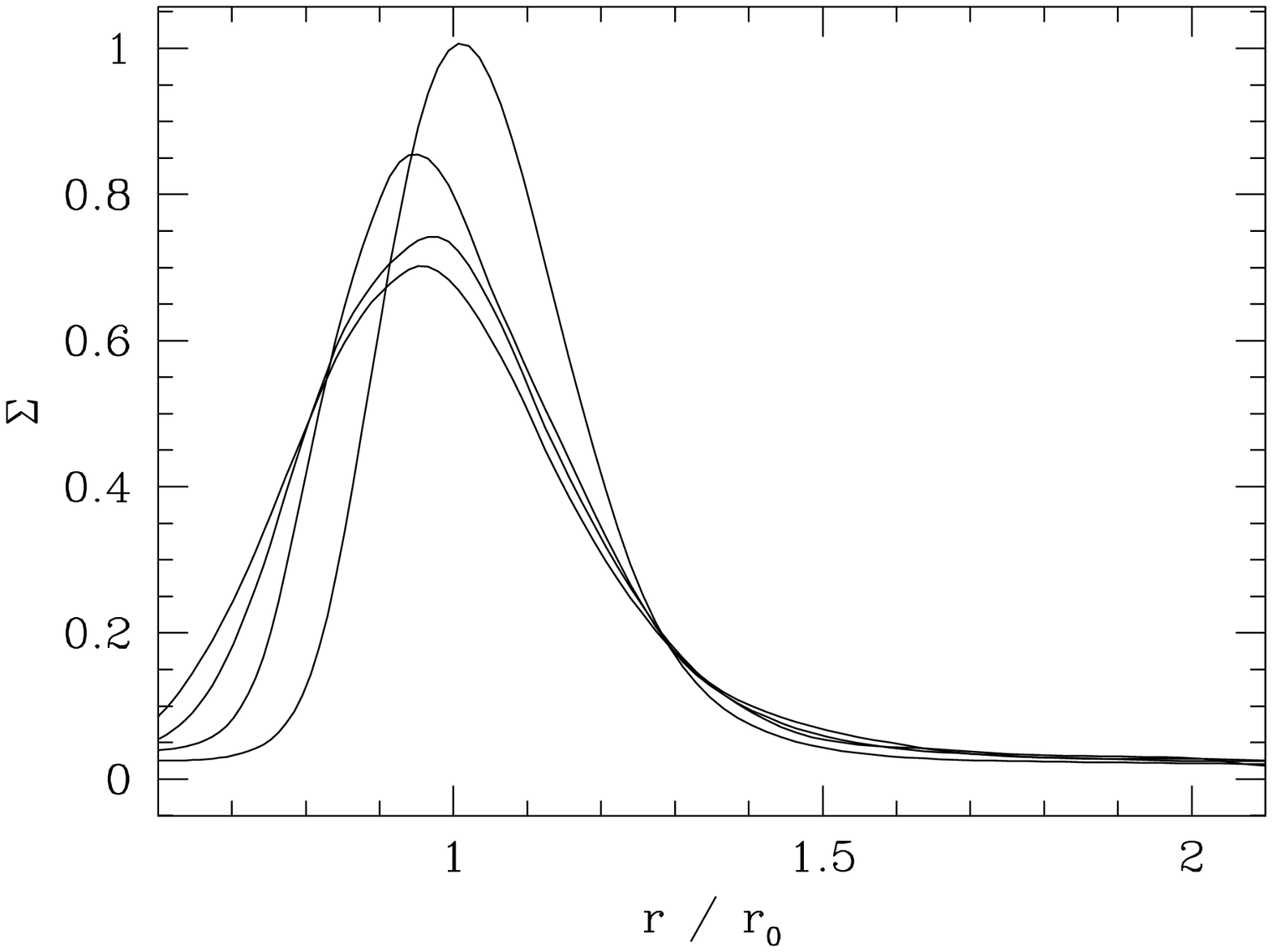}
\vskip-1.3truein
\caption{Evolution of the azimuthally-averaged surface density profile 
(in arbitrary units) from an MHD simulation of the spreading of a narrow 
ring of gas. Curves correspond to the images shown in Fig.~\ref{pja:fig1}
($t = 0$, 2, 4, and 6 orbits at the initial location of the ring). 
The simulated disk profile remains smooth.}
\label{pja:fig2}	
\end{figure}

The second fundamental assumption of one-dimensional models 
is the diffusive evolution of the surface density predicted 
by Equation~\ref{pja:eq_diffuse}. This, too, is broadly 
compatible with what is known about the MRI. 
Figures~\ref{pja:fig1} and \ref{pja:fig2} show the predicted 
evolution of a ring of gas in a Keplerian potential, 
computed directly from a global three-dimensional MHD 
simulation using methods described by 
Armitage (1998). Although large scale non-axisymmetric 
structure is evident in Figure~\ref{pja:fig1}, the 
diffusive evolution of the surface density that is predicted 
by the one-dimensional theory is nonetheless 
reproduced. 

The preceding discussion has emphasized that one-dimensional 
disk models based on the $\alpha$ formalism can provide 
a reasonable description of the evolution of thin disks, 
especially if insights from numerical simulations of the 
MRI are used to guide the detailed modeling.  
Simulations, however, also reveal circumstances under which $\alpha$ 
models fail to describe disks driven by magnetically induced 
turbulence. Papaloizou \& Nelson (2003) and Winters, Balbus, \& 
Hawley (2003a) have shown that averaging the magnetically 
induced stress over tens or even hundreds of orbital times 
is required to yield stable estimates of the average stress. 
In the context of a one-dimensional $\alpha$ disk model, 
this implies that $\alpha$ cannot be regarded as a constant 
over shorter timescales. More generally, any attempt to extend 
$\alpha$ models to treat two-dimensional flows (for example, 
to study eccentric disks or disks strongly 
perturbed by binary companions) is dangerous and not to 
be recommended. Such generalizations frequently rest on the assumption 
that {\em all} the components of the stress tensor in a 
turbulent MHD disk flow behave in the same way as a 
Navier-Stokes viscosity. There is scant reason to believe 
that this is true, and indeed the limited numerical evidence 
collected to date suggests that the multidimensional 
evolution of magnetically driven disks differs significantly 
from that of their viscous counterparts (\cite{pja:nelson03}; 
Winters, Balbus, \& Hawley 2003b).

So far, we have assumed that the disk orbits in the point 
mass potential of the black hole, ignoring the self-gravity 
of the disk itself. This approximation is generally valid 
in the inner regions of the disk and will obviously fail 
if the disk mass becomes comparable to the mass of the black 
hole (perhaps during formation of the hole via accretion). Even 
if the disk mass is relatively modest, however, self-gravity 
can still become important at large enough radii. To demonstrate 
this, we make use of the $\alpha$-prescription and write the 
kinematic viscosity $\nu$ in the form
\begin{equation}
\nu = \alpha { c_s^2 \over \Omega } \,.
\label{pja:eq_alpha} 
\end{equation}  
For a steady disk, away from the boundaries (e.g., \cite{pja:pringle81}), 
\begin{equation} 
 \nu \Sigma = { \dot{M} \over {3 \pi} } \,,
\label{pja:eq_steady}  
\end{equation} 
which, together with Equation~\ref{pja:eq_alpha}, specifies $\Sigma(r)$ 
in terms of the accretion rate, sound speed, and $\alpha$ parameter.
The stability of a nearly Keplerian gas disk against axisymmetric 
perturbations is measured by the Toomre (1964) $Q$ parameter,
\begin{equation}
Q = { {c_s \Omega} \over {\pi G \Sigma} } \,,
\end{equation}
with instabilities developing as $Q \rightarrow 1$ from above. Making use 
of Equations~\ref{pja:eq_alpha} and \ref{pja:eq_steady}, we find that 
for a steady disk,
\begin{equation}
Q \propto { {\alpha c_s^3} \over {\dot{M}} } \,.
\label{pja:eq_Qprofile} 
\end{equation}
Since the temperature and sound speed in the disk are decreasing 
functions of radius (at least if the heating is provided by 
accretion and/or irradiation from the central regions of the 
disk), the outer parts of a steady-state disk 
will inevitably reach the $Q\approx 1$ threshold for self-gravity 
to become important, provided that the disk is large enough. Indeed, 
models of AGN disks suggest that self-gravity typically sets in 
at $r \simeq 10^{-2}$~pc, which is only $\sim 10^3$ Schwarzschild 
radii for a $10^8 \ {\rm M}_\odot$ black hole (\cite{pja:clarke88}; 
\cite{pja:kumar99}; \cite{pja:goodman03}). Interestingly, this 
radius is one to two orders of magnitude {\em smaller} than the radii from 
which maser emission has been observed in NGC4258 (\cite{pja:miyoshi95}) 
and NGC1068 (Greenhill \& Gwinn 1997). This raises the immediate 
question: is the masing disk observed in these galaxies a roughly 
homogenous structure, or is it rather composed of discrete 
clumps (Shlosman et al.\ 1990; \cite{pja:kumar99}; \cite{pja:kartje99})?

%
%
\begin{figure}[tbh]
\includegraphics[width=\textwidth]{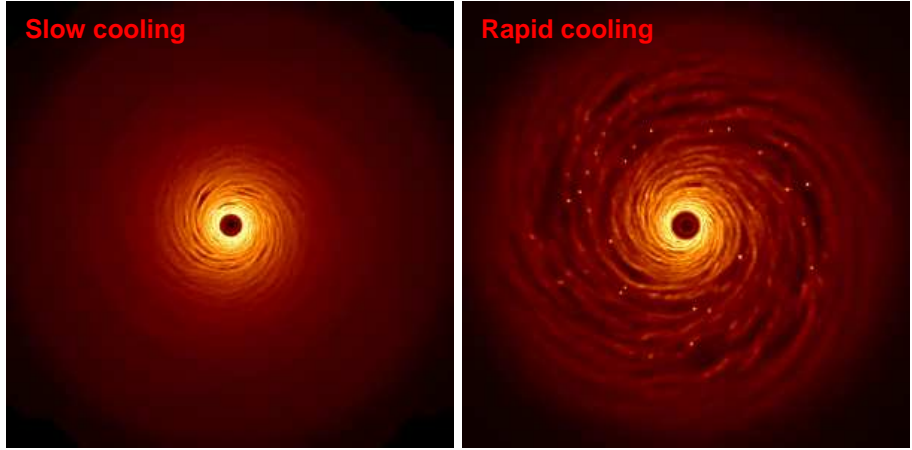}
\vskip-0.1truein
\caption{Surface density from global smooth particle hydrodynamics 
simulations of gravitationally unstable disks. 
The disk in this simulation had a mass 
equal to one tenth of that of the central object. In the 
simulation depicted in the left-hand panel, the 
cooling time $t_{\rm cool} = 5~\Omega^{-1}$. Under these 
conditions, a balance between heating and cooling is 
attainable, and the disk develops a stable spiral 
structure that acts to transport angular momentum 
outward. A shorter cooling time of $t_{\rm cool} = 3~\Omega^{-1}$ 
{\em (right-hand panel)\/} leads to prompt fragmentation of 
the disk into multiple bound objects. 
Note that the thickness of the simulated disk 
exceeded that appropriate for AGN.
Figure adapted from Rice et al.\ (2003; their Figs.~2 and 4).
}
\label{pja:fig3}	
\end{figure}

Once $Q \approx 1$ has been attained, self-gravity triggers the 
development of spiral structure in the disk. Gravitational torques 
then act to transport angular momentum outward, in a manner 
superficially resembling the evolution of an ordinary viscous 
disk described by an $\alpha$ model (\cite{pja:laughlin94}). Such a 
correspondence rests on shaky theoretical foundations (\cite{pja:bp99}), 
because the non-local nature of gravitational angular momentum 
transport immediately violates the local property of $\alpha$ 
models (Eq.~\ref{pja:eq_ss}). That said, examining the 
local properties of self-gravitating disks proves to be 
useful in delineating the circumstances under which a 
self-gravitating disk will fragment, rather than stably 
transport angular momentum (\cite{pja:lodato04}). 
If, at a particular radius, 
the disk can cool via radiative losses on a timescale 
$t_{\rm cool}$, then Gammie (2001) showed that the 
boundary between stable transport and fragmentation 
lies at $t_{\rm cool} = t_{\rm crit} \simeq 3~\Omega^{-1}$. For 
$t_{\rm cool} > t_{\rm crit}$, a stable state can 
be reached in which $Q \sim 1$ and angular momentum 
is transported with an effective $\alpha$ of
\begin{equation}
\alpha = \left[ \left( {9 \over 4} \right) \gamma \left( 
\gamma - 1 \right) \Omega~t_{\rm cool} \right]^{-1} \,,
\end{equation} 
where $\gamma$ is a two-dimensional adiabatic index (\cite{pja:gammie01}). 
As $t_{\rm cool}$ decreases, turbulence in the disk (measured by the 
effective $\alpha$) has to become increasingly violent in order 
to balance the radiative losses, and the overdensities of spiral 
structures increase. Eventually, for $t_{\rm cool} < t_{\rm crit}$, 
the disk fragments into bound objects. Both local (\cite{pja:gammie01}; 
\cite{pja:johnson03}) and global (\cite{pja:rice03}; Fig.~\ref{pja:fig3}) 
simulations broadly support these analytic results.

Application of these ideas to the outer regions of accretion 
disks in AGN is complicated by the dominant role that 
irradiation plays in setting the thermal conditions in 
the outer disk. In particular, fluctuations in the 
disk temperature (and, thus, $Q$) will occur as the 
illumination from the central source varies, with the 
disk being most vulnerable to fragmentation when the 
source is in its lowest state. Furthermore, there is 
no reason to expect that the disk at large radii is 
able to attain the steady-state described by 
Equation~\ref{pja:eq_steady}. 
The most plausible scenario is that the inner non-self-gravitating 
disk is surrounded by a region in which self-gravity transports 
angular momentum in the stable, non-fragmenting regime. Beyond that, 
at radii of order a pc or less, any disk formed from 
inflowing or outflowing gas is unstable to fragmentation 
(\cite{pja:goodman03}). If correct, this suggests that AGN 
need to be resupplied with gas of low specific angular momentum 
(compared to that of a circular orbit at the sphere of 
influence) in order to prevent fragmentation occurring 
in the outer disk. More speculatively, if fragmentation 
does occur, the edge of the accretion disk in AGN could 
be a birth place for massive stars that would ultimately 
form stellar or intermediate mass black holes. If this 
process had occurred in the Galactic Center in the recent 
($\sim$10~Myr) past, it could leave a distinctive signature 
in the form of disk-like kinematics for some subset of 
massive stars (\cite{pja:levin03}). Moreover, if some fraction 
of the stars ended up embedded in the inner disk, they 
would migrate inward on at most a viscous timescale. 
Gravitational waves from the final inspiral of the 
stellar remnants with the supermassive black hole 
are potentially detectable with the {\em Laser Interferometer
Space Antenna} ({\em LISA}; \cite{pja:levin04}; 
\cite{pja:tan04}).

In broad outline then, the MRI is likely to provide the 
main mechanism of angular momentum transport in the inner 
disk, while self-gravity is potentially important at 
relatively large radii. 
A more subtle question is whether there is a region of the 
disk in which both processes are operative, or, 
conversely, a zone in which neither drives efficient 
angular momentum transport. This depends upon whether 
the MRI is able to operate in the cool outer regions 
of the disk, where the gas has a low ionization fraction 
and the flux-freezing assumption of ideal MHD breaks 
down. For AGN, a quantitative discussion of the 
effect of the low temperature on angular momentum 
transport processes has been given by Menou \& Quataert (2001). 
For a disk with sound speed $c_s$ and resistivity $\eta$, 
the magnetic Reynolds number is
\begin{equation}
 {\rm Re}_M = { {v_A h} \over \eta } \,.
\label{pja:eq_rem} 
\end{equation} 
If ${\rm Re}_M$ is too low, then diffusion of magnetic fields 
due to the finite conductivity can suppress the MRI 
(e.g., \cite{pja:matsumoto95}; \cite{pja:gammie96}; 
\cite{pja:wardle99}; \cite{pja:sano02}). 
The critical value of ${\rm Re}_M$ (which may also need to take 
account of the Hall effect) is probably 
of order unity or somewhat larger. Identical reasoning 
suggests that ambipolar diffusion can also shut off the MRI 
and prevent MHD disk turbulence from developing (\cite{pja:menou01}). 

Applying these ideas to AGN disks, Menou \& Quataert (2001) 
found that the ionization state of the outer disk---and hence 
the likelihood of the MRI being operative---was highly uncertain, 
depending upon (among other things) whether the outer disk was 
exposed to irradiation from the central source. If the MRI does 
shut off at radii where $Q \gg 1$, then 
an intermediate zone in the disk could be both magnetically and 
gravitationally inactive. Such a ``dead zone'' would act to 
starve the inner disk of resupply from larger radii, until 
enough matter built up to initiate self-gravity 
or magnetorotational activity. Complicated and poorly 
understood time-dependent accretion would probably 
result (e.g., \cite{pja:gammie99b}; \cite{pja:alp01}). 
Alternatively, if the regions of MHD turbulence and 
gravitational instability overlap, then there could be 
non-trivial interactions between the two angular momentum 
transport mechanisms. Planet formation simulations 
(Winters et al.\ 2003b) show that {\em external} gravitational 
torques can modify the properties of MHD disk turbulence, and 
the first simulations of self-gravitating magnetized disks 
(\cite{pja:fromang04}) likewise show that magnetic and 
gravitational torques are not simply additive. In the 
calculations of Fromang et al.\ (2004), the presence of 
MHD turbulence in a self-gravitating disk significantly 
reduced the rate of gravitational transport of angular momentum.

Finally, it remains possible that accretion is not always 
driven by angular momentum redistribution 
within the disk at all, but rather by angular momentum loss 
in an outflow. Outflows and jets are certainly present in some 
AGN, and if magnetically driven (\cite{pja:blandford82}), they 
can potentially remove {\em all} of the angular momentum needed 
to allow accretion, while ejecting only a modest fraction of 
the mass. Models in which outflows are entirely responsible 
for driving accretion are not popular, in part because of 
skepticism that the required magnetically launched outflows 
are present across a sufficiently broad range of disk radii. 
In particular, although there is near-universal agreement that 
magnetic fields are implicated in the formation of relativistic 
jets, the slower outflows that originate further away from 
the black hole may well be partially or entirely accelerated 
by radiation pressure acting on resonance lines (\cite{pja:vitello88}; 
\cite{pja:murray95}; \cite{pja:proga03}). Moreover, the stability 
of accretion arising from magnetic outflow torques is rather  
doubtful (\cite{pja:lubow94}; \cite{pja:cao02}; but see 
also \cite{pja:konigl96} and \cite{pja:campbell01}). Nonetheless, 
the possibility that angular momentum loss via MHD outflows 
dominates over internal stresses in restricted regions of 
AGN disks remains entirely viable. It constitutes a significant 
source of uncertainty in our understanding of the structure of 
the disk and when considering possible mechanisms for 
large-scale variability.

\section{The Effect of the Accretion Rate}
\label{secpjaaccrate}

The accretion rate onto a black hole is most usefully measured 
as a fraction of the accretion rate at which radiation pressure
becomes important for the dynamics of the flow. For a
spherically symmetric flow, radiation pressure acting 
on free electrons balances gravity at the Eddington 
luminosity,
\begin{equation} 
 L_{\rm Edd} = { {4 \pi G m_p c} \over {\sigma_T} } M \,,
\end{equation} 
where $\sigma_T$ is the Thomson scattering cross-section.
If the luminosity is generated by accretion with radiative 
efficiency $\epsilon$, then $L = \epsilon \dot{M} c^2$ and the 
above expression defines an Eddington accretion rate. 
Taking $\epsilon = 0.1$, 
\begin{equation} 
 \dot{M}_{\rm Edd} = 2.2 \left( { M \over {10^8 \ {\rm M}_\odot}} \right) \ 
 {\rm M}_\odot~{\rm yr}^{-1} \,.
\end{equation} 
The accretion rate can then be expressed in dimensionless form 
as a fraction $\dot{m} \equiv \dot{M} / \dot{M}_{\rm Edd}$ 
of the Eddington rate. Current theoretical 
understanding identifies $\dot{m}$ as the most important 
parameter controlling the structure and observational 
appearance of black hole accretion flows. We will consider 
in turn the three cases: $\dot{m} \ll 1$, $\dot{m} \sim 1$, 
and $\dot{m} \gg 1$. These cases, by definition, differ in 
the importance of radiation pressure for the dynamics of the 
flow. Equally important, the ability of the flow to cool 
also varies systematically with $\dot{m}$.

For $\dot{m} \ll 1$, the critical question is whether 
the inflowing gas can radiate the gravitational potential 
energy liberated by accretion. Most theoretical work 
suggests that the vast majority of the energy dissipated 
by MHD turbulence in a disk goes initially into heating 
the ions rather than the electrons (\cite{pja:eliot98}; 
\cite{pja:blackman99}; \cite{pja:eliot99}; \cite{pja:medvedev00}), 
at least as long as the magnetic field remains sub-equipartition. 
Sub-equipartition fields are the 
normal outcome of the non-linear phase of the MRI in MHD simulations 
of accretion flows in the fluid limit (for a discussion of how the MRI 
itself is altered if the plasma is collisionless, see 
\cite{pja:eliot02}). Radiative losses from bremsstrahlung radiation, 
synchrotron radiation, and Comptonization of soft photons, on the 
other hand, primarily take energy from the electrons. In this 
situation, where the ions are heated simultaneously with the 
electrons being cooled, the extent of coupling between the 
two fluids obviously determines the state of the plasma. If 
the dominant coupling process is ordinary Coulomb collisions, 
then the timescale for thermal equilibrium to be established 
between the ions and the electrons is (\cite{pja:spitzer62}; 
\cite{pja:ichimaru77})
\begin{equation} 
\tau_{ei} = { 
{ 3 \left( kT_e \right)^{3/2} m_p } \over 
{ 8 n e^4 \sqrt{ 2 \pi m_e} \ln \Lambda } } \,.
\end{equation}  
Here, $T_e$ is the electron temperature, $k$ is the Boltzmann 
constant, $m_p$ is the proton (ion) mass, 
$n$ is the number density of ions, and $m_e$ is the electron mass. The 
Coulomb logarithm has a value of $\ln \Lambda \sim 20$. When 
the accretion rate is low, $n$ is small, and this, coupled with the 
high temperature near the black hole (of the ions generally, and 
the electrons, as well, if we initially postulate a one-temperature 
plasma), implies a 
large value for $\tau_{ei}$. Once $\tau_{ei}$ becomes comparable 
or larger than the inflow time, electrons and ions can no 
longer maintain equilibrium via Coulomb collisions. Instead,
a two-temperature plasma develops in which the ions stay 
close to the virial temperature while the electrons are 
substantially cooler. All current models for low $\dot{m}$ 
accretion flows onto black holes are based on 
the two-temperature idea (e.g., \cite{pja:ichimaru77}; \cite{pja:rees82}; 
\cite{pja:narayan94}; \cite{pja:abramowicz95}; \cite{pja:blandford99}; 
and the numerous variants on these models), and would not be 
viable otherwise. In particular, any collective mechanism 
that coupled the ions to the electrons substantially more 
efficiently than Coulomb collisions would undercut the 
foundations of two-temperature plasma accretion models. 
Although no such mechanism is known, the well known 
complexities of laboratory plasmas suggest that the 
possible existence of more efficient coupling cannot 
be entirely ruled out (\cite{pja:begelman88}; \cite{pja:bk97}; 
\cite{pja:eliot98}; \cite{pja:eliot99}; \cite{pja:medvedev00}).

Although from a purely theoretical perspective the question 
of whether low $\dot{m}$ accretion flows form a two-temperature 
plasma remains open, observations---particularly those that 
indicate a remarkably low luminosity for the black hole at 
the Galactic Center (\cite{pja:narayan98})---are 
hard to reconcile with almost any imaginable single temperature 
disk model (see below). This has prompted most theorists to accept 
that accretion flows with $\dot{M} \ll \dot{M}_{\rm Edd}$ develop a 
two-temperature structure, though the exact threshold below 
which a hot two-temperature flow supplants a cool thin disk 
is less certain. Narayan \& Yi (1995b) suggest that the critical 
accretion rate is around
\begin{equation}
 \dot{m}_{\rm crit} \simeq \alpha^2 \,;
\label{pja:eq_adaf} 
\end{equation}  
that is, one to two orders of magnitude below the Eddington limit.

\begin{figure}
\includegraphics[width=\textwidth]{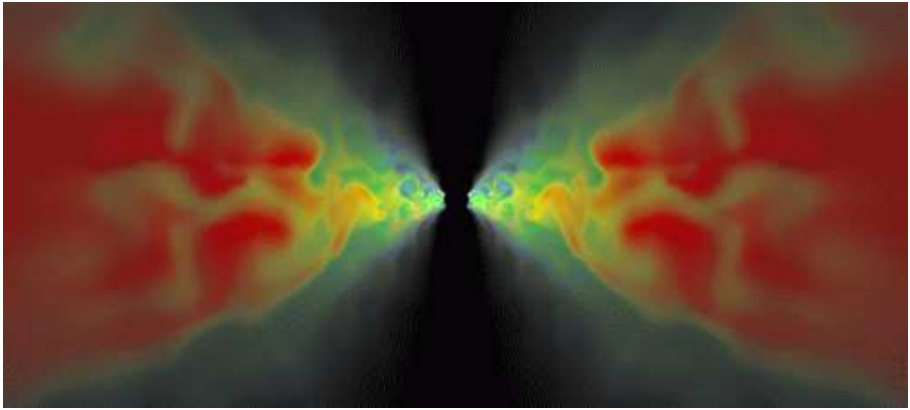}
\vskip-0.1truein
\caption{Structure of a radiatively inefficient accretion flow, 
from unpublished axisymmetric simulations by the author and 
Kees Dullemond. The numerical methodology is identical to 
that used by Stone, Pringle, \& Begelman (1999). Such flows 
are geometrically thick, turbulent, and weakly bound. Outflows 
can readily be driven off the surface of the disk and along the polar 
funnel regions.}
\label{pja:fig4}	
\end{figure}

Once a two-temperature flow has developed, three model-independent 
conclusions follow. First, since the ion temperature $T_i$ remains 
comparable to the virial temperature, the flow can be {\em 
geometrically thick}, with $h/r \sim 1$. Second, since most 
of the accretion energy remains locked up in the ions, the 
flow is {\em underluminous}, with an accretion efficiency 
much smaller than the $\epsilon \simeq 0.1$ values realized for 
thin disks. Finally, since the gas is both hot and rapidly 
rotating, it is at most marginally bound to the black hole 
(formally, the Bernoulli constant, which is conserved 
for adiabatic flows, is positive).
As a consequence, it is certainly plausible, 
and possibly inevitable, that a substantial fraction of the putative 
fuel for the black hole actually ends up outflowing from the galactic 
nucleus (\cite{pja:narayan95a}; \cite{pja:blandford99}, 2004).
Outflows further reduce the already low radiative efficiency 
(now defined as $\epsilon = L / c^2 \dot{M}$, where the inflow rate is 
measured at large radius), because most of the gas then fails 
to sample the deep potential well close to the black hole. 
Figure~\ref{pja:fig4} 
shows semi-quantitatively (not quantitatively, because the simulation 
depicted was two-dimensional and non-magnetic) what the resulting flow 
looks like.

%
%

Apart from such generalities, no consensus has yet been reached 
as to the detailed structure of non-radiative accretion flows 
(also described as advection dominated flows, since most of 
the liberated gravitational potential energy is advected with 
the fluid rather than being radiated). The problem lies in the 
fact that the structure of 
non-radiative accretion flows is expected to depend 
sensitively on the nature of energy and angular momentum 
transport. The numerical simulations needed to directly attack 
this issue are extraordinarily difficult. For an illustration 
of the difficulties, 
suppose that the black hole is accreting from the hot interstellar 
medium of an elliptical galaxy with $T_{\rm ISM} \sim 10^6 \ {\rm K}$ 
and sound speed $c_s$ a few hundred kilometers per second. The 
ratio of the Bondi radius,
\begin{equation} 
 R_B = { {GM} \over c_s^2 } \,, 
\end{equation}  
to the gravitational radius of the hole---both of which 
must be resolved if we want to get the boundary conditions 
strictly correct---is then
$R_B / R_g \sim 10^6$, vastly exceeding the dynamical 
range accessible to any present three-dimensional 
simulation. Two-dimensional simulations can do better
but have their own limitations due to both the antidynamo 
theorem in axisymmetry (\cite{pja:cowling34}) and the fact 
that the MRI has different properties 
in axisymmetry, as compared to three dimensions 
(\cite{pja:goodman94}; Hawley et al.\ 1995). Recent 
examples of simulations include work by Stone \& Pringle (2001),
Machida, Matsumoto, \& Mineshige (2001), 
Hawley \& Balbus (2002), Proga \& Begelman (2003), Pen, Matzner, \& 
Wong (2003), and Igumenshchev, Narayan, \& Abramowicz (2003). There is 
substantial agreement that outflows of various kinds occur 
generically as a side effect of non-radiative accretion, but 
opinions still differ on the steady-state radial structure 
of the flow, and on how to interpret the numerical results 
in terms of simpler analytic models.

%
%
\begin{figure}
\includegraphics[width=\textwidth]{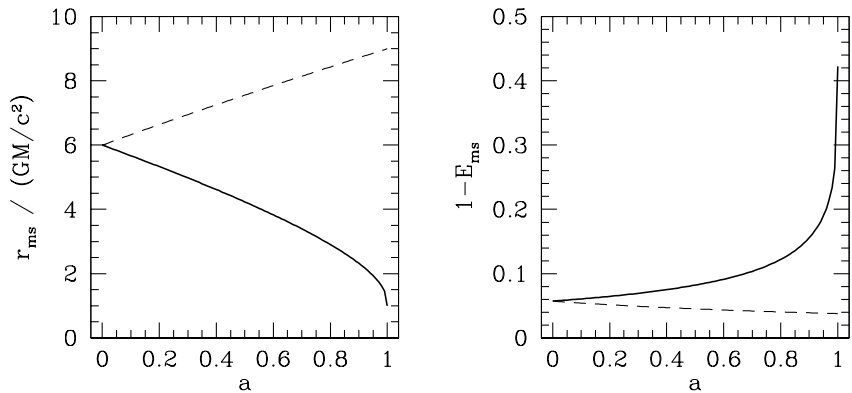}
\vskip-2.5truein
\caption{{\em (Left panel)} Location of the marginally stable 
circular orbit as a function of dimensionless spin parameter $a$
for prograde {\em (solid line)} and retrograde {\em (dashed line)} 
orbits. {\em (Right panel)} Efficiency of energy extraction if 
gas spirals slowly into $r_{\rm ms}$ before plunging silently 
into the black hole. Plotted is $1-E_{\rm ms}$, where 
$E_{\rm ms}$ is the binding energy of the marginally stable 
circular orbit for prograde and retrograde motion.}
\label{pja:fig5}	
\end{figure}

Accretion rates above $\dot{m}_{\rm crit}$ (Eq.~\ref{pja:eq_adaf}) 
yield flows that are dense enough to provide good coupling between 
ions and electrons. Provided that the accretion rate is not 
extremely high, i.e., $\dot{m} \sim 1$, the flow is then radiatively 
efficient and cools to form a geometrically thin ($h/r \ll 1$) 
disk. The disk extends from an outer radius, which is probably 
set by the considerations of gravitational stability described 
in the previous section, down to the marginally stable circular 
orbit at $r = r_{\rm ms}$. Interior to the marginally stable 
circular orbit, which lies at $r_{\rm ms} = 6~GM / c^2$ for a 
Schwarzschild black hole, circular particle orbits are unstable---any 
inward radial peturbation suffices to place the particle onto a 
plunging trajectory into the black hole. 
In the simple---but possibly incorrect---case where there is zero 
torque at $r_{\rm ms}$ (non-zero torque is 
discussed in \S\ref{secpjadyn}), 
the radiative efficiency is determined by the binding energy $E_{\rm ms}$ 
of gas at that radius. Plots of these quantities are shown as 
a function of the dimensionless spin parameter ($a \equiv Jc / GM^2$ 
for a hole with an angular momentum $J$) in Figure~\ref{pja:fig5}.

The basic theory of thin disks is well known (for reviews, see, e.g., 
\cite{pja:pringle81}; Frank et al.\ 2002). The main open issues 
(at least for flat, circular disks) concern the role of magnetic fields 
and radiation. The structure of magnetic fields 
near $r_{\rm ms}$ is of particular interest, since it 
determines not only the boundary condition at the disk's 
inner edge (\cite{pja:krolik99}), but also the strength 
of magnetic fields threading the black hole itself. If the 
black hole is spinning, magnetic fields admit the possibility 
of tapping some of the rotational energy via the 
Blandford-Znajek (1977) process. Spin energy extraction 
provides one way to power jets (e.g., \cite{pja:blandford00}), 
and is frequently suggested---mostly on the basis of empirical 
arguments---as the source of the radio-loud/radio-quiet 
dichotomy in AGN (e.g., \cite{pja:wilson95}). As a counter 
to such claims, Ghosh \& Abramowicz (1997) and 
Livio, Ogilvie, \& Pringle (1999) have argued that since 
dynamo generated disk fields are expected to be weak 
(sub-equipartition), and dragging in ordered external 
fields is difficult (Lubow et al.\ 1994), the efficiency 
of the Blandford-Znajek process is likely to be negligibly 
small for thin disks. These arguments are plausible, 
but I doubt that they represent the last word on the subject, 
not least because of observational indications for spin 
energy extraction in some AGN (\cite{pja:wilms01}; \cite{pja:reynolds04}, 
though note that even if spin energy {\em is} being extracted, this 
need not be the classical Blandford-Znajek process at work).

As the accretion rate increases, thin disks are predicted to 
remain the preferred mode of accretion below some upper limit 
set by the effects of radiation pressure. For a spherical 
flow, that upper limit lies at $\dot{m} = 1$. In disks, 
however, matters need not be so simple. 
Once radiation pressure dominates gas pressure in the 
inner disk, the disk is prone to several instabilities, 
even in the absence of magnetic fields (\cite{pja:ss76}; 
\cite{pja:agol01}). When magnetic fields {\em are} included, 
it is found that the flow develops photon 
bubbles (\cite{pja:arons92}; \cite{pja:gammie98a}; 
\cite{pja:blaes01}, 2003)---low density radiation filled 
cavities that coexist with dense gas pressure dominated 
regions. In principle, radiation can escape from such a 
highly inhomogenous structure at rates above the Eddington 
limiting luminosity, {\em without} being accompanied by 
catastrophic mass loss (\cite{pja:begelman01}; \cite{pja:begelman02}; 
see also \cite{pja:shaviv98}). This could permit the 
existence of stable, thin disks, with an emergent luminosity  
as large as $L=10-100~L_{\rm Edd}$ 
(\cite{pja:begelman02}). Detailed calculations of the 
outcome of these instabilities in magnetized disks, and 
their implications for the structure of radiation 
pressure dominated flows even below the Eddington limit, 
are now in progress (\cite{pja:turner02}; \cite{pja:turner03}; 
\cite{pja:turner04}).

The final regime to consider is that of overfed accretion, with 
$\dot{m} \gg 1$. By definition, radiation pressure is central 
to the dynamics of such flows, though it need not act to 
shut off accretion directly. At sufficiently high accretion 
rates, the opacity can be large enough that photons in 
the inner regions are advected inward faster than they 
can random walk away from the black hole. Eventually, they 
are dragged across the event horizon along with the gas and
are not radiated from the system. For a spherical accretion 
flow, it is straightforward to show that advection wins 
over diffusion inside a trapping radius given by
\begin{equation}
r_{\rm trap} = { {\dot{M} c^2} \over L_{\rm Edd} } 
{ r_s \over 2} \,,
\end{equation}
where $r_s = 2GM / c^2$ is the Schwarzschild radius 
(\cite{pja:begelman79}). The same behavior occurs 
for flows with non-zero angular momentum (\cite{pja:begelman82}). 
The emergent luminosity from photon-trapped flows remains 
around $L_{\rm Edd}$ (e.g., \cite{pja:blondin86}; \cite{pja:ohsuga02}), 
so as $\dot{m}$ increases, the radiative efficiency decreases 
inversely with the accretion rate. 

Clearly, the similarities between the $\dot{m} \ll 1$ and the 
$\dot{m} \gg 1$ regimes are striking. In both cases, the 
radiative efficiency is low, for $\dot{m} \ll 1$ because 
of inefficient ion-electron coupling, and for $\dot{m} \gg 1$ 
because the high opacity prevents photons from escaping the 
fluid. In the simplest analysis, overfed accretion will 
then also result in a hot, geometrically thick disk, with 
the main difference from the underfed case being the lower 
value of $\gamma$ appropriate for a radiation dominated gas. 
Strong outflows could then be inevitable, and these might 
even reduce the accretion rate onto the hole back down to  
levels comparable to the Eddington limit. Once again, simulations of 
radiation dominated disks will be needed to 
test such speculations.

In addition to providing an explanation for the dichotomy in the 
observed properties of supermassive black holes in AGN and 
quiescent galactic nuclei, the existence of these different 
modes of accretion also has implications 
for the growth of black holes. Accretion rates well below the 
Eddington limiting value probably provide the black hole 
with a negligible amount of mass, despite the fact that 
this mode lasts much longer than more active phases 
(\cite{pja:priya98}). More surprisingly, attempts to 
flood the black hole with mass at rates vastly in excess 
of the Eddington value may also fail if, as we have 
speculated, such high $\dot{m}$ flows are as vulnerable 
to outflows as their low $\dot{m}$ counterparts. 
The entire accretional growth of black holes may be 
restricted to occur via radiatively efficient accretion at rates that lie 
within an order of magnitude or so of the Eddington 
limit (probably extending {\em above} as well as below the 
nominal ``limit''). For the 
most luminous AGN, such a ``what you see is what you get'' 
model allows for a consistent match (within substantial 
uncertainties) between observations of the luminous output 
of AGN and the local density of supermassive black holes 
(\cite{pja:amy01}; \cite{pja:merritt01}; \cite{pja:yu02}; 
following the basic approach of \cite{pja:soltan82}). It 
may also apply to the earlier growth phase of supermassive 
black holes.

\section{Dynamics of the Plunging Region}
\label{secpjadyn}

%
%
\begin{figure}[t]
\includegraphics[width=\textwidth]{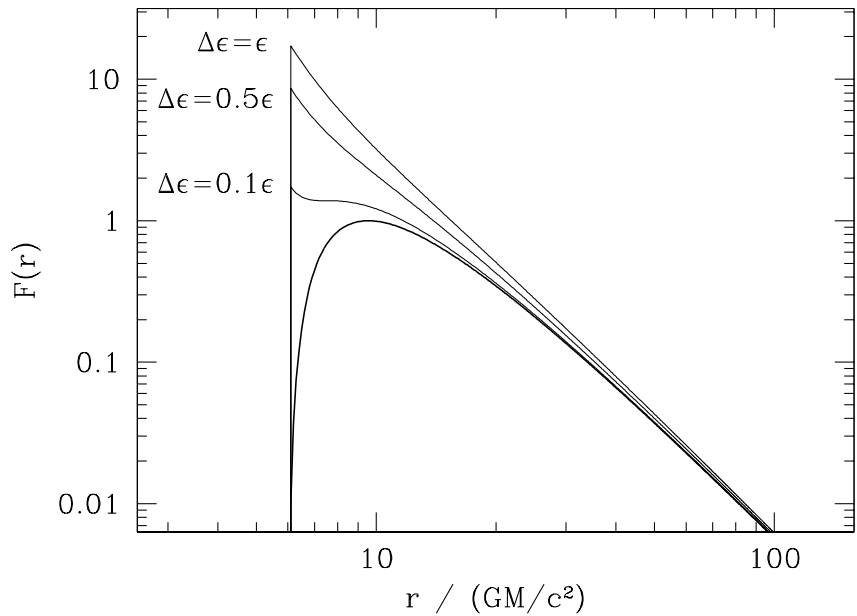}
\vskip-1.3truein
\caption{Local flux from the disk (i.e., the emission per unit area
in the rest frame of the orbiting gas, ignoring returning radiation)
for disk models with
different inner boundary conditions. Lowest curve shows
the profile of the emission for a standard Novikov-Thorne
disk (Page \& Thorne 1974) around a Schwarzschild black hole,
assuming zero torque at $r_{\rm ms}$. $F(r)$ peaks well outside
the marginally stable circular orbit. The three upper curves
show the effect of adding increasing amounts of torque at
$r_{\rm ms}$ (Agol \& Krolik 2000), with the labels showing the
resulting increase $\Delta \epsilon$ in the radiative efficiency as
a fraction of the original efficiency $\epsilon$. No emission is
assumed to originate from the plunging region.}
\label{pja:fig6}
\end{figure}

The inner boundary of a radiatively efficient accretion disk 
around a black hole lies close to the marginally stable 
circular orbit. At $r > r_{\rm ms}$, the force balance for a 
fluid element in a thin disk is predominantly between gravity 
and centrifugal force, with angular momentum transport driving 
a slow, subsonic inflow. Interior to $r_{\rm ms}$, circular 
orbits are unstable. In the standard model for black hole accretion 
disks (\cite{pja:novikov73}; \cite{pja:page74}), once gas 
reaches this plunging region, it flows into the black hole along 
geodesics, maintaining a constant energy and angular momentum. 
In this simple model, gas in the plunging region can have 
observational effects, mainly by reprocessing or reflecting 
radiation generated elsewhere in the system (\cite{pja:cunningham76}; 
\cite{pja:reynolds97a}). The plunging region does not, however, 
alter the structure of the disk or the radiative efficiency of the 
flow. The inner boundary condition for the disk is simply that 
there should be zero torque at $r=r_{\rm ms}$.

Early studies of black hole accretion assumed (often explicitly) 
that magnetic fields in the plunging region remained weak. Provided 
that this is so, the inflowing gas rapidly attains supersonic 
(and super-Alfv\'enic) radial velocities and becomes causally 
disconnected from the disk outside the marginally stable 
orbit. This justifies the zero-torque boundary condition at 
$r_{\rm ms}$. If magnetic fields become strong, for example 
as a consequence of the shearing of frozen-in fields during the 
final inspiral, the argument fails. Energy and angular momentum 
can then be exported from the plunging region to the disk, modifying 
its structure and emission (\cite{pja:krolik99}; \cite{pja:gammie99a}). 
The same basic process can occur in either Schwarzschild or Kerr 
geometry.

The consequences of a non-zero torque at the marginally stable orbit 
have been discussed in detail by Agol \& Krolik (2000). The 
most basic is that energy that would otherwise be lost down the 
hole can be transmitted from the plunging region to the disk, and 
subsequently radiated. As a result, the radiative efficiency of 
thin disk accretion could be larger than normally assumed 
(Fig.~\ref{pja:fig5}). 
In particular, the standard value of $\epsilon = 0.06$ for the yield 
of accretion onto Schwarzschild holes may more usefully be regarded 
as a lower limit. This implies that---although the qualitative trend 
for more rapidly spinning black holes to be more luminous at a given 
accretion rate than Schwarzschild holes is very reasonable---there 
are significant theoretical uncertainties that need to be borne 
in mind when attempting to determine the average spin of black 
holes from comparison of black hole masses and integrated AGN 
output. For example, the relatively high inferred radiative efficiency 
($\epsilon > 0.1$) derived by Elvis, Risaliti, \& Zamorani (2002)
and Yu \& Lu (2004) need not necessarily imply a 
rapid average spin for the supermassive black hole 
population. Perhaps more interestingly, given 
that measuring $\epsilon$ to a factor of about two precision is 
obviously a challenging task, a non-zero torque at the disk inner 
edge makes a large change to the radial distribution of the disk 
emissivity. This is plotted in Figure~\ref{pja:fig6}, using 
expressions given in Agol \& Krolik (2000). Unlike a ``standard'' 
disk, in which the local flux peaks well outside the last 
stable orbit, a torqued disk has a steeply rising emissivity 
profile right down to the marginally stable orbit (at large 
radius, the contribution to the emissivity due to the torque 
at $r_{\rm ms}$ scales with radius as $F(r) \propto r^{-7/2}$, which 
is steeper than the dissipation profile for an untorqued disk). This 
difference in the dissipation with radius is potentially 
observable. It is worth noting that similarly steep 
dissipation profiles are predicted in some models in 
which magnetic fields act to extract energy from a 
Kerr black hole (Takahashi et al.\ 1990; \cite{pja:li02}).

%
%
\begin{figure}
\includegraphics[width=\textwidth]{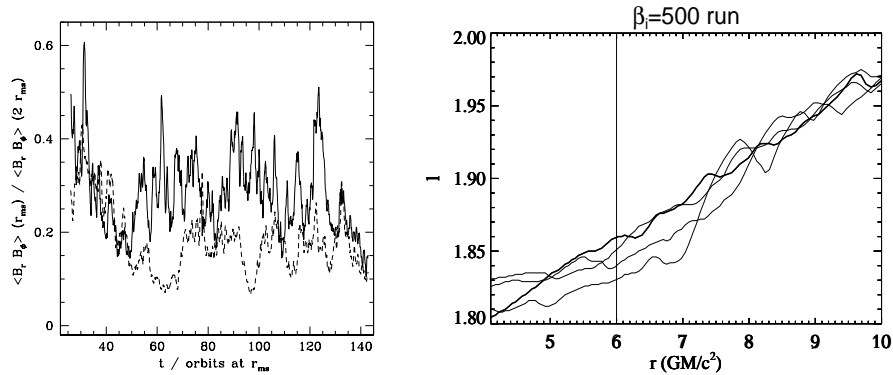}
\vskip-0.1truein
\caption{{\em (Left panel)} Magnetic torque at the marginally stable
orbit as a fraction of the torque in the disk at a radius
of $2~r_{\rm ms}$, as derived from numerical simulations
with relatively high {\em (solid line)} and relatively low
{\em (dashed line)} sound speed (Armitage \& Reynolds 2003).
{\em (Right panel)} Specific angular momentum $l(r)$ at several
different times for a simulated flow in which the disk magnetic
field strength was relatively large (Reynolds \& Armitage 2001).
The inner disk is quite variable, but typically $l(r)$
continues to decline interior to $r_{\rm ms}$.}
\label{pja:fig7}
\end{figure}

Numerical simulations have generally been supportive of the basic 
concept outlined by Krolik (1999). Pseudo-Newtonian calculations, 
which use a modified potential to mimic some of the most important 
relativistic effects (\cite{pja:paczynski80}), have confirmed that 
there can be significant non-zero torques at $r_{\rm ms}$ 
(\cite{pja:hawley00}; \cite{pja:reynolds01}; Hawley \& Krolik
2001, 2002).
As shown in Figure~\ref{pja:fig7}, this leads to a declining 
specific angular momentum profile that extends seamlessly 
through the last stable orbit to smaller radii. The same 
phenomenon has now been seen in General Relativistic 
MHD simulations of black hole accretion (\cite{pja:devilliers03}; 
\cite{pja:devilliers03b}). 
Large amplitude (factor $\sim 2$) fluctuations in the torque also 
occur and persist over a timescale of $\sim 10$ orbital periods. 
The importance of this torque for the dynamics of the disk 
may also vary with the temperature (or, equivalently, thickness) 
of the inner disk, with the zero-torque limit becoming a 
better approximation as the thickness decreases 
(\cite{pja:armitage01}; \cite{pja:reynolds01}; 
\cite{pja:armitage03}; \cite{pja:afshordi03}). 

\section{Variability}

The emission from AGN is variable (\cite{pja:mushotzky93}; 
\cite{pja:ulrich97}), as is the much weaker emission from the 
Galactic Center (\cite{pja:baganoff01}; \cite{pja:zhao01}; 
\cite{pja:genzel03}; \cite{pja:ghez04}). Although in some 
observationally important cases the {\em correlations} between 
variability in different wavebands can be understood through 
simple light travel time arguments (\cite{pja:blandford82b}), 
the basic origin of most of this variability is not well 
understood. Here, I discuss what insights into the problem recent MHD 
calculations of disk accretion provide.

The first question we might ask is, are there 
large amplitude changes in the luminosities of AGN akin to 
those seen in most other disk accreting systems? We 
have already discussed the outbursts of dwarf novae, which 
are attributed to a global thermal instability of the 
accretion disk. Accretion onto some neutron stars and 
(stellar mass) black holes in low-mass X-ray binaries 
is similarly prone to thermal instability 
(\cite{pja:tanaka96}) and can be modeled using a 
minimally modified version of the theory developed 
for dwarf novae (\cite{pja:dubus01}). The same ideas 
have also been pressed into service to explain the 
outbursts of protostellar disks seen in  
FU Orionis objects (\cite{pja:hartmann96}; \cite{pja:bell94}), 
though the identification of these events with thermal 
disk instabilities is less compelling.

If thermal instabilities act on a global scale in AGN, 
the timescales for the cycling between outburst and 
quiescence would be long---probably $\sim 10^4 \ {\rm yr}$ 
or longer (\cite{pja:lin86}). Transitions between ``on'' and 
``off'' states might not occur on easily observable timescales, 
though they could leave traces in the form of intermittent 
activity from radio-loud AGN (\cite{pja:reynolds97b}; 
\cite{pja:owsianki98}). If such intermittent 
accretion is commonplace, there ought to exist a population of 
relatively quiescent galaxies  
in which the black hole is surrounded not by a radiatively 
inefficient flow, but rather by a temporarily inactive thin 
disk. Detailed models of this type have been developed by 
Siemiginowska, Czerny, \& Kostyunin (1996) 
and by Siemiginowska \& Elvis (1997).

Disks in AGN extend out to radii that encompass the zone in 
which hydrogen is partially ionized, so the basic ingredient 
necessary for thermal instabilities to occur is present 
(\cite{pja:andy98}). What is less clear is whether the 
efficiency of angular momentum transport (i.e., the value 
of $\alpha$) varies between the ionized and neutral state 
in the AGN environment. A substantial difference 
between $\alpha_{\rm hot}$ and $\alpha_{\rm cold}$ is needed 
if the whole disk is to flip between outburst and 
quiescent states. In dwarf novae, the required change 
probably occurs because the hot and cold states lie 
on opposite sides of the critical degree of ionization 
required for sustained MHD turbulence (Eq.~\ref{pja:eq_rem};
\cite{pja:gammie98b}). It has been argued (\cite{pja:menou01}) 
that this is no longer the case in AGN disks, leading to 
thermal instabilities that produce local flickering but 
not global outbursts. If so, any observational evidence 
for large scale outbursts in AGN may require a different 
explanation. Numerous possibilities for time dependent 
behavior suggest themselves in the outer disk, where 
there is an interplay of gravitational and MHD disk 
instabilities, though none have been worked out in any 
great detail.

At frequencies between $\sim 10^{-4} \ {\rm Hz}$ and 
$\sim 10^{-8} \ {\rm Hz}$ 
(i.e., timescales of tens of minutes to around a year), X-ray 
observations provide good constraints on the properties of 
AGN variability (e.g., \cite{pja:edelson99}; \cite{pja:uttley02}; 
\cite{pja:vaughan03}; \cite{pja:markowitz03}; and references therein).  
The power spectral density $G(f)$ derived from these observations 
is often consistent with a power law broken at a frequency 
$f_{\rm break}$,
\begin{eqnarray} 
 G(f) & \propto &  f^{-\alpha_{\rm low}} \,\,\,\,\, f < f_{\rm break} 
\nonumber \,, \\
 G(f) & \propto & f^{-\alpha_{\rm high}} \,\,\,\, f > f_{\rm break} \,. 
\end{eqnarray} 
Representative values 
for the two slopes are $\alpha_{\rm low} \simeq 1$ and 
$\alpha_{\rm high} \ge 2$ (Vaughan \& Fabian 2003; 
\cite{pja:markowitz03}).

%
%
\begin{figure}[t]
\includegraphics[width=\textwidth]{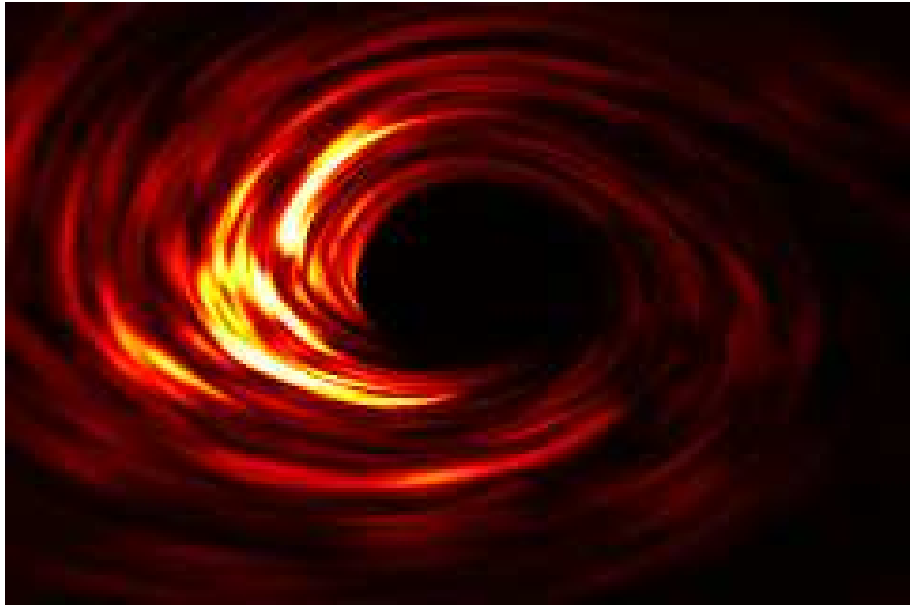}
\vskip-0.0truein
\caption{View of a simulated disk as seen by a distant observer at
an inclination angle of 55$^\circ$ (Armitage \& Reynolds 2003).
The patchy emission is boosted on the approaching
(left hand) side of the disk as a consequence of beaming.
In constructing this image, the local disk emission has been
assumed to scale with the vertically integrated magnetic stress.
Raytracing through the Schwarzschild metric is then used to account
for the relevant relativistic effects, including beaming,
gravitational redshift, and light bending.}
\label{pja:fig8}
\end{figure}

\begin{figure}[t]
\includegraphics[width=\textwidth]{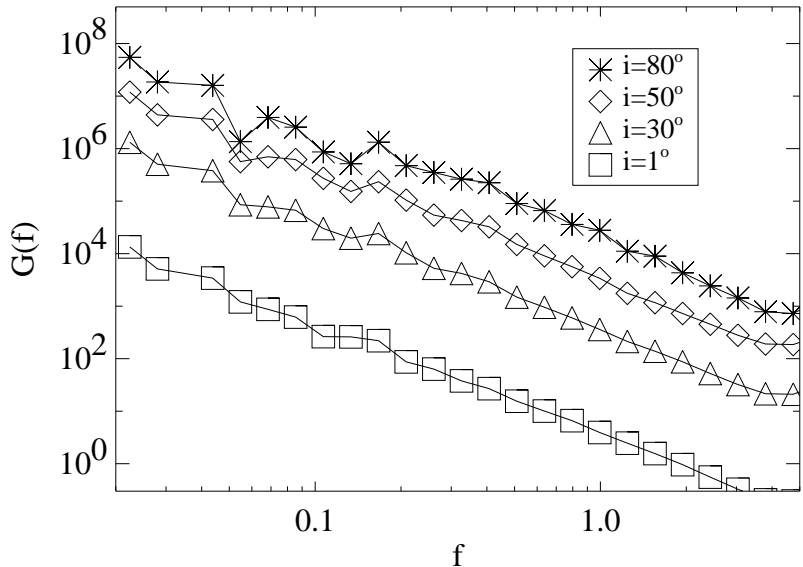}
\vskip-0.2truein
\caption{Temporal power spectra $G(f)$ derived from a numerical 
simulation of accretion onto Schwarzschild black holes 
(Armitage \& Reynolds 2003). The frequency $f$ has been rescaled 
such that the orbital frequency at the radius of marginal 
stability corresponds to $f=1$.  Shown here are the power spectra 
for $i=1^\circ$, $i=30^\circ$ (offset by $10^2$), 
$i=50^\circ$ (offset by $10^3$),
and $i=80^\circ$ (offset by $10^4$). MHD turbulence in the disk 
generates variability across a wide range of timescales, with 
quantities such as the predicted emission or mass accretion rate 
showing steep, approximately power law spectra.}
\label{pja:fig9}	
\end{figure}

Although a detailed comparison is beyond the capabilities of current 
numerical simulations, there appear to be good prospects for relating 
noise power spectra of this form to theoretical models of MHD disk 
turbulence. Simulations of magnetically active disks show that 
the flow displays variability across a wide range of timescales, 
with physical quantities such as the magnetic stress and mass 
accretion rate showing temporal power spectra that are described 
by power laws (\cite{pja:kawaguchi00}; \cite{pja:hk02}; \cite{pja:armitage03}). 
The patchy and rapidly fluctuating pattern of stress across the surface 
of the inner disk, which gives rise to this variability, is 
shown in Figure~\ref{pja:fig8}. If we assume 
that the local disk emission traces the magnetic stress, then 
the power spectra predicted from the simulations have the 
form shown in Figure~\ref{pja:fig9}. Regardless of the 
inclination of the disk to the line of sight (which affects 
the importance of relativistic effects such as beaming), 
the temporal power spectrum at frequencies comparable 
to those of the inner disk is around $G(f) \propto f^{-2}$. 
Individual annuli in the simulated disk, moreover, produce 
power spectra that break---at around the orbital frequency---to 
substantially steeper slopes of around $f^{-3.5}$ 
(\cite{pja:armitage03}). These results support the 
common assumption that the break frequency in observed 
systems scales with the orbital frequency at the 
marginally stable orbit, and hence with the mass of 
the black hole (e.g., \cite{pja:nowak00}; \cite{pja:lee00}). 
Although there are significant uncertainties in many 
AGN black hole masses, the observations of Markowitz et al.\ (2003) 
suggest a correlation between black hole mass and 
break frequency. We note, however, that work to date has not 
been able to recover the observed values of 
$\alpha_{\rm low}$ and $\alpha_{\rm high}$---the theoretical 
power spectra are steeper than observed.
%
%

\section{Warped, Twisted, and Eccentric Disks}

Flat disks, in which the gas follows close to circular orbits, 
are special cases of the more general situation in which the 
disk may be eccentric and/or warped out of a single plane. 
A common assumption is that an initially warped or eccentric 
disk will relax to a flat circular one ``on a viscous timescale''. 
This is at best a gross simplification of the evolution of 
warped or eccentric disks, and may even be qualitatively wrong. 
As one might expect, the evolution of warped or eccentric disks 
is rather complex, and in this section I attempt only to 
provide a framework for understanding some of the key results. 
For more details, the interested reader is well advised 
to consult the recent and exhaustive studies 
by Ogilvie (1999, 2000, 2001), which include references to the 
earlier literature on the subject. 

Observations provide clear motivation for considering the 
evolution of warped disks in AGN. The masing disks in 
NGC4258 (\cite{pja:miyoshi95}), NGC1068 (\cite{pja:greenhill97}), 
and the Circinus galaxy (Greenhill et al.\ 2003) all appear to  
be warped, albeit to different degrees. This raises two 
questions. First, are the warps merely decaying features that 
reflect the complex angular momentum distribution of the infalling 
gas that formed the disk, or are they self-excited by some 
process intrinsic to the disk or AGN itself? Second, if the 
warps are in fact decaying, how rapidly does that 
process occur compared to the timescale on which the 
gas would be accreted? 

Although easily formulated, both of these questions have proved 
to be difficult to answer. The basic physics revolves around 
the fact that in a warped disk, neighboring annuli are inclined 
relative to one another, and as a consequence, there 
is shear in the vertical direction as well as in the radial 
direction. Dissipation of the energy associated with the 
out-of-plane motions is governed by the $(r,z)$ component 
of the stress, and this acts to flatten out the disk as, 
simultaneously, the $(r,\phi)$ stress drives accretion. 
The crucial point is that these stresses can be generated by 
entirely different physical processes (for a concrete example, 
see \cite{pja:gammie00})
and need not act on the same timescale, even roughly 
(\cite{pja:pringle92}). The detailed physics of stresses 
within the disk will then determine whether the disk 
flattens before a significant fraction of the gas 
accretes, or whether instead the warp is advected inward 
along with the gas.

To go further, we need to elaborate on the mechanisms that 
can allow annuli to exchange angular momentum. If the 
disk is thin enough, specifically, if
\begin{equation}
 {h \over r} < \alpha \,,
\end{equation} 
then the evolution of the tilt as a function of radius, 
like the evolution of the surface density, is diffusive 
(\cite{pja:papaloizou83}). If, conversely,
\begin{equation}
 {h \over r} > \alpha \,,
\end{equation} 
then the warp is communicated through the disk by waves 
(\cite{pja:papaloizou95}). Assuming that 
$\alpha \sim 0.1$ in geometrically thin AGN disks, then
most parts of the disk (except, perhaps, the 
innermost regions) are likely to fall into the 
diffusive regime, which we will consider exclusively 
from now on. We will also ignore the complications of self-gravity 
(Papaloizou, Terquem, \& Lin 1998), even though, as already noted, 
the masing disks cited as motivation are probably 
self-gravitating.

In the diffusive limit, the evolution of the disk can 
be described to a first approximation using the simple 
theory described by Pringle (1992). In this theory, 
the stresses within the disk are reduced to two 
viscosities; $\nu_1$, which describes the usual 
kinematic viscosity leading to inflow, and $\nu_2$, 
which is related to the stresses which act to flatten 
out any warp. The timescales for accretion $t_\nu$ and 
for decay of the warp $t_{\rm warp}$ are related to 
these viscosities in the usual way:
\begin{eqnarray}
t_\nu & = & { r^2 \over \nu_1 } \,, \\ \nonumber 
t_{\rm warp} & = & { r^2 \over \nu_2 } \,.
\end{eqnarray} 
The relation between these two viscosities can be 
calculated either analytically (\cite{pja:ogilvie99})
or via numerical simulations (\cite{pja:ulf00}). Both 
approaches suggest that
\begin{equation} 
{ \nu_2 \over \nu_1 } \simeq { 1 \over {2 \alpha^2} } \gg 1 \,.
\end{equation}
This is a striking result. It implies that warps are 
likely to decay rapidly and will require strong forcing 
if they are to persist for long periods. Torques 
from the reprocessing of radiation from the central 
source by the disk provide one well studied forcing 
term, which could act to excite warps even in 
initially planar disks (\cite{pja:pringle96}; 
\cite{pja:maloney96}). The large ratio of 
$\nu_2 / \nu_1$ implied by the above equation, 
however, severely limits the circumstances in 
which radiation driven warping can overcome the 
disk's intrinsic tendency to flatten out. Torques 
from a disk wind are potentially more promising 
(\cite{pja:schandl94}; \cite{pja:lai03}), though harder to 
evaluate in detail.

Further complications ensue if 
the black hole is rotating, even modestly. Orbits 
whose angular momentum is inclined to the spin axis 
of a rotating hole will precess due to the Lense-Thirring 
effect (``frame dragging''). This forced precession 
is only significant at small radii, where it will 
affect the inclination of the disk. Again, the 
simplest case to analyze is the one in which the 
inner disk is in the diffusive regime of warp 
evolution. Differential precession, acting on 
an initially flat but misaligned disk, will 
rapidly induce a sharp twist in the disk 
close to the marginally stable orbit. This 
will be flattened out by viscosity, 
with the consequence that the disk near 
a spinning hole will be driven into the equatorial 
plane of the hole, regardless of the inclination 
at large radii (Bardeen \& Petterson 1975). 
Over longer timescales, 
the torque between the disk and the hole will also 
act to realign the system until the angular 
momentum vector of the hole, and of the disk 
at {\em all} radii, are coincident. Since 
normally $J_{\rm disk} \gg J_{\bullet}$, this 
balancing act will usually result in a large 
change in the spin axis of the black hole
until it roughly matches the angular momentum 
vector of gas in the outer disk. How quickly 
this process occurs depends, again, largely 
on $\nu_2 / \nu_1$, with large 
values of this ratio leading to rapid alignment 
(\cite{pja:priya98b}).

In some circumstances, the inner accretion disk 
around a rotating black hole may be thick enough 
that warps display wave-like evolution. In this 
case, the inner disk may be able to remain misaligned 
with respect to the spin axis of the black hole, even 
in the presence of dissipation (Demianski \& Ivanov 1997;
\cite{pja:ivanov97}; \cite{pja:lubow02}). This could 
have important consequences for the direction of jets 
launched from the inner regions of the disk.

Unlike in the case of warped disks, there is little 
observational evidence to suggest that AGN disks---or any disk 
that is not strongly forced, for example by a binary 
companion---are significantly 
eccentric, though some sort of asymmetry is 
deduced in as many as 60\% of AGN with 
double-peaked Balmer lines (\cite{pja:strateva04}). 
Theoretical studies, however, raise the interesting possibility 
that there could be circumstances in which isolated accretion disks 
spontaneously develop eccentricity. In particular, 
a two-dimensional disk model in which the viscosity 
follows a Navier-Stokes form, with shear viscosity 
$\mu$ and bulk viscosity $\mu_b$, is unstable to the 
growth of eccentricity if
\begin{equation}
 { {{\rm d}\ln \mu} \over {{\rm d}\ln \Sigma} } > 
 1 + {1 \over 3} \left( { \mu_b \over \mu } - {2 \over 3} \right) \,;
\end{equation}
that is, if the vertically integrated viscosity is increasing 
too rapidly with surface density (\cite{pja:ogilvie01}). 
If the bulk viscosity is small, as is often assumed in 
accretion disk modeling, then this result implies that 
almost all disks ought to be unstable to the growth 
of eccentricity (\cite{pja:postnov94}; \cite{pja:ogilvie01}).

Earlier, of course, we specifically warned 
of the dangers of assuming that angular momentum transport 
in an MHD flow can be described via an effective 
Navier-Stokes viscosity. Indeed, Ogilvie (2001) shows 
that if angular momentum transport follows an alternative 
analytic prescription, which is motivated by the 
phenomenology of MHD turbulence, then the wholesale 
instability of disks to eccentricity can be avoided. 
The important point is that the behavior of eccentric disks---
which at a minimum will exist in galactic nuclei 
following the tidal disruption of stars (\cite{pja:rees88}),  
or when binary black holes are present---is closely 
related to subtle aspects of turbulent transport within 
the disk. Predictions as to their evolution 
cannot yet be made with confidence.

\begin{acknowledgments}
I'm grateful to Chris Reynolds for reading and commenting on 
an earlier draft of this Chapter.
\end{acknowledgments}

\end{document}